\documentclass[pss]{wiley2sp} % provides new 2008 pss two-column style (no alternative manuscript style output available at present)

\usepackage{amsmath}
\begin{document}

% Title of the article

\title{The method of the likelihood and the Fisher
information in the construction of physical models}

\titlerunning{Likelihood method and Fisher information in construction of physical models}

\author{%
  E.W. Piotrowski\textsuperscript{\textsf{\bfseries 1}},
  J. S{\l}adkowski\textsuperscript{\textsf{\bfseries 2}},
  J. Syska\textsuperscript{\textsf{\bfseries 2,\Ast}},
  S. Zaj{\c a}c\textsuperscript{\textsf{\bfseries 2}}
  }

% Abbreviated list of authors for the page haeders
\authorrunning{Piotrowski, S{\l}adkowski, Syska, Zaj{\c a}c}

%E-mail-address of corresponding author
\mail{\ e-mail
  \textsf{jacek.syska@us.edu.pl}, Phone
  +xx-xx-xxxxxxx, Fax +xx-xx-xxx}

% author's affiliations/addresses
\institute{%
    \textsuperscript{1}\,Institute of Mathematics, The University of
Bia{\l}ystok, Pl-15424 Bia{\l}ystok, Poland\\
\textsuperscript{2}\,Institute of Physics, University of Silesia,
Uniwersytecka 4, 40-007 Katowice, Poland}

\received{XXXX, revised XXXX, accepted XXXX} % do not change, will be filled in by the publisher
\published{XXXX} % do not change, will be filled in by the publisher

%Please select four to six PACS-codes from the enclosed list (PACS.txt) or from www.aip.org/pacs)
 % For example: 71.20.Ps

\abstract{%
% This is a macro for the typesetting of two-column text in an
% abstract. It will typeset the two arguments in \abstcol{}{} as the
% left and right column inside the abstract box. At the
% columnbreak there will be always a columnbreak (\par), so both
% columns start with a new paragraph. No automatic column height
% balancing is done.
%
% If used with a \titlefigure it will silently output both
% parameters as consecutive paragraphs.
%
% The macro is defined exclusively inside the argument of \abstract{};
% if used outside it will raise an error.
%
% Usage: \abstcol{<left column>}{<right column>}
%\abstcol{%
The subjects of the paper are the likelihood method (LM) and the
expected Fisher information ($FI$) considered from the point od
view of the construction of the physical models which originate in
the statistical description of phenomena. The master equation case
and structural information principle are derived. Then, the
phenomenological description of the information transfer is
presented. The extreme physical information (EPI) method is
reviewed. As if marginal, the statistical interpretation of the
amplitude of the system is given. The formalism developed in this
paper would be also applied in quantum information processing and
quantum game theory.
%  }
  }

% The class file requires the standard graphicx Latex package. See the 'LaTeX
% standard graphics and color packages documentation' for more information at
% <http://tug.ctan.org/tex-archive/macros/latex/required/graphics/grfguide.pdf>.
%
% Accepted figure file formats depend on which LaTeX flavour is used.
% Classic LaTeX is always able to use Encapsulted Postscript (EPS);
% PDFLaTeX can't use this but accepts PDF, JPG, PNG, and GIF formats.
%
% See examples for implementing graphics in floating figure environments later in this file.
% If \titlefigure is given, it takes as its mandatory parameter the
% name (without extension) of some figure file.
%\titlefigure[height=3.1cm]{empty2w}
%\titlefigurecaption{%
% This is the caption of the \emph{optional} abstract figure. If
%  there is no figure here, the abstract text will fill both columns.}

\maketitle   % please do not remove

\section{Introduction.}

The subjects of the paper are the likelihood method (LM) and the
expected Fisher information ($FI$), $I$, considered from the point
of view of the construction of the physical models, which
originate in the statistical description of phenomena. The $FI$
had been introduced by Fisher \cite{Fisher} as a part of the new
technique of parameter estimation of models which undergo the
statistical investigation in the language of the maximization of
the value of the likelihood function $L$ (signed below as $P$).
The notion of the $FI$
%(or channel capacity for the Stam's information case)
is also connected with the Cram{\'e}r-Rao inequality, being the
{\it maximal} inverse of the possible value of the mean-square
error of the unbiased estimate from the true value of the
parameter. The set $\Theta = (\theta_{n})$ of the
parameters\footnote{In the most general case of the estimation
procedure the dimensions of $\Theta \equiv (\theta_{i})_{1}^{k}$
and the sample ${\bf y} \equiv ({\bf y}_{n})_{1}^{N}$ are usually
different, yet because $(\theta_{i})$ is below the set of physical
"positions" of a physical nature (e.g. positions in the case of an
equation of motion or energies in the case of a statistical
physics generating equation \cite{Frieden}) we have $E({\bf
y}_{n}) = \theta_{n}$ and the dimensions of $\Theta$ and ${\bf y}$
are the same, i.e. $k=N$.} composes the components in the
statistical space ${\cal S}$. Under the regularity conditions the
expected $FI$ (here because of the summation over $n$, the {\it
channel capacity}):
%defined as:
\begin{eqnarray}  \label{Fisher_information_1}
I &\equiv&   \sum_{n=1}^{N}  \int \! d^{N} {\bf y} \frac{-
\partial^{2} ln P( {\bf y}; \, \Theta)}{\partial \theta_{n}^{2}}
P({\bf y}) \nonumber \\ &=&  \sum_{n=1}^{N}  \int \! d^{N} {\bf y}
( \frac{\partial ln P( {\bf y}; \, \Theta)}{\partial \theta_{n}}
)^{2} P({\bf y})
\end{eqnarray}
is the variance of the score function $\mathrm{S}(\Theta) \equiv
\partial ln P(\Theta)/\partial \Theta$ \cite{Pawitan}. It
describes the local properties of $P( {\bf y}; \; \theta_{1},...,
\theta_{N} )$ which is the joint probability (density) of the $N$
data values ${\bf y} \equiv ({\bf y}_{1}, ..., {\bf y}_{N})$,  but
is understood as a function of the parameters $\theta_{n}$. Below
with ${\bf x}_{n} \equiv {\bf y}_{n} - \theta_{n}$ we note the
added fluctuations of the data ${\bf y}_{n}$ from $\theta_{n}$.

The development of the statistical methods introduced by Fisher
\cite{Fisher} has followed along two different routes. The first
one, of the differential geometry origin, began when C.R. Rao
\cite{Rao} noticed that the $FI$ matrix defines a Riemannian
metric. For a short review of the following work, until the
consistent definition of the $\alpha$-connection on the
statistical spaces see \cite{Amari Nagaoka book}. The second one
is based on the construction of the information (entropical)
principles and we will describe this approach in details. It was
put forward by Frieden and his coauthors, especially Soffer
\cite{Frieden}, and is known under the notion of the extreme
physical information~(EPI). Frieden began the construction of the
physical models by obtaining the kinetic term from the
$FI$\footnote{Let for a system the square distance between two of
its states denoted by $q({\bf x})$ and $q'({\bf x}) = (q + d q)
({\bf x})$ could be written in the Euclidean form $\; d s^{2} =
\int_{X} d {\bf x} \,dq \, dq \;$ where $X$ is the space of the
positions {\bf x} (in one measurement). Supposing that the states
are described by the probability distributions we compare it with
the distance $\; d s^{2} = \frac{1}{4} \int_{X} d {\bf x} \frac{dp
\, dp}{p({\bf x})} \;$ in the space ${\cal S}$ with the Fisher-Rao
metric \cite{Bengtsson_Zyczkowski}, where $p({\bf x})$ and $p' =
(p + dp)({\bf x})$ are the probability distributions for these two
states. Then we notice that these two formulas for $d s^{2}$
coincide if $q({\bf x}) = \sqrt{p({\bf x})}$. {\bf In this way the
notion of the {\it amplitude} $\, q({\bf x})$ appears naturally
from the Riemannian, Fisher-Rao metric of the space ${\cal S}$.}
It satisfies the condition: $\, \int_{X} \, d {\bf x} \, p({\bf
x}) = \int_{X} \, d {\bf x} \, q({\bf x}) \,q({\bf x}) = 1 \;$.

Now, let the sample of dimension $N$ be "collected" by the system
and $q_{n}$ are the real amplitudes and $p_{x_{n}} ({\bf x}_{n})$
are the probability distributions with the property of the {\it
shift invariance} $p_{x_{n}} ({\bf x}_{n})$ $= p_{x_{n}} ({\bf
x}_{n}|\theta_{n}) = p_{n} ({\bf y}_{n}|\theta_{n})$ where ${\bf
x}_{n} \equiv {\bf y}_{n} - \theta_{n}$. Assuming that the data
are collected independently which gives the {\it factorization
property} $P({\bf y}) \equiv P({\bf y}|\Theta) = \prod_{n=1}^{N}
p_{n}({\bf y}_{n}|\theta_{n})$, where $\theta_{m}$ has no
influence on ${\bf y}_{n}$ for $m \neq n$, and using the chain
rule $\partial/\partial \theta_{n} = \partial/\partial ({\bf
y_{n}} - \theta_{n}) \; \partial ({\bf y_{n}} -
\theta_{n})/\partial \theta_{n} = - \; \partial/\partial ({\bf
y_{n}} - \theta_{n}) = - \; \partial/\partial {\bf x_{n}}$, the
transition  from the statistical form of $FI$ given by
Eq.(\ref{Fisher_information_1}) to its kinematical form given
below by the first possibility:
\begin{eqnarray}  \label{Fisher_information}
I \! = \! 4 \! \sum_{n=1}^{N} \int \!\! d {\bf x}_{n} \! \left(
\frac{\partial q_{n}}{\partial {\bf x}_{n}} \right)^{2}  {\rm or}
\;\, 4  N \! \sum_{n=1}^{N/2} \int \!\! d{\bf x}  \frac{\partial
\psi_{n}^{*}({\bf x})}{\partial {\bf x}} \frac{\partial
\psi_{n}({\bf x})}{\partial {\bf x}}
\end{eqnarray}
is performed \cite{Frieden}. The second possibility is achieved
\cite{Frieden} as follows: Using the amplitudes $q_{n}$, the wave
function $\psi_{n}$ could be constructed as $ \psi_{n} =
\frac{1}{\sqrt{N}} (q_{2 n - 1} + i \, q_{2 n})$, where $n = 1, 2,
..., N/2 $ with the number of the real degrees of freedom being
twice the complex ones. With this, the first kinematical form of
$I$ in Eq.(\ref{Fisher_information}) could be rewritten in the
second form, where additionally the index $n$ has been dropped
from the integral as the range of all ${\bf x}_{n}$ is the same.
Finally, the total probability law for all data gives $p({\bf x})$
$= \sum_{n=1}^{N} p_{x_{n}} ({\bf x}_{n}|\theta_{n}) P(\theta_{n})
= \frac{1}{N} \sum_{n=1}^{N} q_{n}^{2}$, where we have chosen
$P(\theta_{n}) = \frac{1}{N}$ \cite{Frieden}. All of these lead to
$p({\bf x}) = \sum_{n=1}^{N/2} \psi_{n}^{*} \psi_{n}$. With the
shift invariance condition and the factorization property, $I$
does not depend on the parameter set $(\theta_{n})$
\cite{Pawitan}, and the wave functions $\psi_{n}$ do not depend on
these parameters, i.e. (expected) positions, also \cite{Frieden}.
%hence we have $p({\bf x})=|\psi({\bf x})|^2$ rather than $|\psi({\bf x}|\Theta)|^2$.
From this summary of the Frieden method we notice that although
the wave functions $\psi_{n}$ are complex the quantum mechanics is
really the statistical method.}. \\
Then he postulated the variational (scalar) information principle
(IP) and internal (structural) one. Each of the IP has the
universal form, yet its realization depends on the particular
phenomenon. The variational IP leads to the dispersion relation
proper for the system. The structural IP describes the inner
characteristics of the system and enabled \cite{Frieden} the
fixing of the relation between the $FI$ (the channel capacity) $I$
and the {\it structural information} ($SI$) $Q$ \cite{Syska}. The
interesting point is that a lot of calculations is performed when
the so called physical information $K$ is partitioned equally into
$I$ and $Q$ (or with the factor $1/2$), having the total value
equal to zero. The method is fruitful as Frieden derived (i) the
Klein-Gordon eq. for the field with the rank $N$ (for the scalar
field $N=2$), (ii) the Dirac eq. for the spinorial field with
$N=8$, (iii) the Rarita-Schwinger eq. for higher $N$ and (iv) the
Maxwell eqs. for $N=4$. The generality of the method enables to
describe the general relativity theory and take into account gauge
symmetry also \cite{Frieden}. Using the information principles
(IPs), Frieden gave the information theory based method of
obtaining the Heisenberg principle which from the statistical
point of view is the kind of the Cram{\'e}r-Rao theorem with the
$FI$ of the system written, after the Fourier transformation, in
the momentum representation\footnote{In reality the Fourier
transformation forms the type of the entanglement between the
variables of the system.}. Then the formalism enabled both to
obtain the upper bound for the rate of the changing of the entropy
of the analyzed system
and to derive the classical statistical physics also \cite{Frieden}. \\
In our paper we develop the less informational approach to the
construction of the IP using the notion of the {\it total physical
information} ($TPI$), $K=Q+I \,$ \cite{Syska}, instead of the
change of the physical information $K=I-J$ introduced in
\cite{Frieden}. This difference does not affect the derivation of
the equation of motion (or generating eq.) for the problems
analyzed until now \cite{Frieden}, yet the change of the notion of
$K$ to $TPI$ with its partition into the kinematical and
structural degrees of freedom, goes closely with the recent
attempts of the construction of the principle of entropy
(information) partition. In Section~\ref{structural principle} we
will derive the structural IP from the analysis of the Taylor
expansion for the log-likelihood function. Our approach to the IP
results also in the change of the understanding of the essence of
the information (entropy) transfer in the process of measurement
when this transfer from the structural to the kinematical degrees
of freedom takes place. We will discuss it in
Section~\ref{information transfer}. Despite the differences we
still call this approach the Frieden one and the method the EPI
estimation. Finally, with the interpretation of $I$ as the
kinematical term of the theory, the statistical proof on the
impossibility of the derivation of the wave mechanics and field
theories for which the rank $N$ of the field is finite from the
classical mechanics, was given \cite{Syska}.

\section{The master equations vs structural information principle}
\label{physical estim} The maximum likelihood method (MLM) is
based on the set of $N$ likelihood equations \cite{Fisher}:
\begin{eqnarray}
\label{MLM basic} \mathrm{S}(\Theta)\mid_{\Theta = \hat{\Theta}}
\; \equiv \frac{\partial}{\partial \Theta} ln P(\Theta)
\mid_{\Theta = \hat{\Theta}} = 0 \; ,
\end{eqnarray}
where the MLM's set of $N$ estimators $\hat{\Theta} \equiv
(\hat{\theta}_{n})_{1}^{N}$ is its solution. These $N$ conditions
for the estimates maximize the likelihood of the sample.

\subsection{The master equations.}
\label{master eq} Yet, we could approach to the estimation
differently. So, after the Taylor expanding of $P(\hat{\Theta})$
around the true value of $\Theta$ and integrating over the sample
space we obtain:
\begin{eqnarray}
\label{L expand} & & \!\!\!\!\!\!\!\! \int \!\! d^{N}{\bf x} \!
\left( P(\hat{\Theta}) - P(\Theta) \right)  =  \!\! \int \!\!
d^{N}{\bf x} \! \left(  \sum_{n=1}^{N} \! \frac{\partial
P(\Theta)}{\partial \theta_{n}} (
\hat{\theta}_{n} - \theta_{n} ) \right. \nonumber \\
&+& \left. \frac{1}{2} \sum_{n,n'=1}^{N} \frac{\partial^{2}
P(\Theta)}{\partial \theta_{n'} \partial \theta_{n}} (
\hat{\theta}_{n} - \theta_{n} ) (  \hat{\theta}_{n'}  -
\theta_{n'} ) + \cdots \right)
\end{eqnarray}
where the notation $P(\Theta) \equiv P(\hat{\Theta})\mid_{\Theta}$
has been used. When the integration over the whole sample space is
performed then neglecting the higher order terms\footnote{Which is
exact even at the density level if only $P(\hat{\Theta})$ $\in$
${\cal S}$ has not higher than the 2-order jets at $\Theta$.} and
using the normalization condition $ \int \, d^{N}{\bf x} \,
P(\Theta) = \int \, d^{N}{\bf x} \, P(\hat{\Theta}) = 1$ we see
that the LHS of Eq.(\ref{L expand}) is equal to zero. Hence we
obtain the result which for the
%distributions \cite{Pawitan},
locally unbiased estimators \cite{Amari Nagaoka book} and after
postulating the zeroing of the integrand at the RHS, takes for
particular $n$ and $n'$ the following microscopic form of the {\it
master equation}:
\begin{eqnarray}
\label{L expand microscop} \frac{\partial^{2} P(\Theta)}{\partial
\theta_{n'} \partial \theta_{n}} ( \hat{\theta}_{n} - \theta_{n} )
(  \hat{\theta}_{n'}  - \theta_{n'} ) = 0 \; .
\end{eqnarray}
When the parameter $\theta_{n}^{\nu}$ has the Minkowskian index
$\nu$ then $P = \prod_{n=1}^{N} p_{n}(x_{n}^{\nu})$ and Eq.(\ref{L
expand microscop}) leads, after the transition to the Fisherian
variables (as in Footnote~2), to the form of the equation of
conservation of flow:
\begin{eqnarray}
\label{conservation flow eq} \frac{\partial
p_{n}(x_{n}^{\mu})}{\partial t_{n}} + \sum_{i=1}^{3}
\frac{\partial \, p_{n}(x_{n}^{\mu})}{\partial x_{n}^{i}} \,
\hat{v}_{n}^{i} = 0 \; ,  \;\;\; \hat{v}_{n}^{i} \equiv
\frac{\hat{\theta}_{n}^{i} - \theta_{n}^{i}}{\hat{\theta}_{n}^{0}
- \theta_{n}^{0}} \; ,
\end{eqnarray}
where $t_{n} \equiv x_{n}^{0}$. Here $\theta_{n}^{i}$ and
$\theta_{n}^{0}$ are the expected position and time of the system,
respectively and index $n$ could be omitted.

\subsection{The information principle. The EPI method.}
\label{structural principle}

Using $ln P$ instead of $P$ and after the Taylor expanding of $ln
P(\hat{\Theta})$ around the true value of $\Theta$ and integrating
with the $d^{N}{\bf x} \, P(\Theta)$ measure over the sample space
we obtain (instead of Eq.(\ref{L expand})) the equation of the EPI
method of the model estimation:
\begin{eqnarray}
\label{Freiden like equation} & & \!\!\! \int \!\! d^{N}{\bf x}
P(\Theta) ( ln \frac{P(\hat{\Theta})}{ P(\Theta)} - R_{3} - \!
\sum_{n=1}^{N} \frac{\partial ln P(\Theta)}{\partial \theta_{n}} (
\hat{\theta}_{n} - \theta_{n} ) ) \nonumber
\\ & & \!\!\!\! = \! \frac{1}{2} \! \int \!\! d^{N}{\bf x} \, P(\Theta) \!\!\!\! \sum_{n,n'=1}^{N} \!\!\!
\frac{\partial^{2} ln P(\Theta)}{\partial \theta_{n'}
\partial \theta_{n}} ( \hat{\theta}_{n} - \theta_{n} ) ( \hat{\theta}_{n'} -
\theta_{n'} )
\end{eqnarray}
with $P(\Theta) \equiv P(\hat{\Theta})\mid_{\Theta}$.  The term on
the LHS of Eq.(\ref{Freiden like equation}) has the form of the
modified relative entropy. Let us define the (observed) structure
$\texttt{t\!F}$ of the system as follows:
\begin{eqnarray}
\label{structure T} \texttt{t\!F} \equiv  ln
\frac{P(\hat{\Theta})}{ P(\Theta)} - R_{3} \; .
\end{eqnarray}
%and when Eq.(\ref{MLM basic}) is fulfilled,  Eqs.(\ref{L expand})
%and (\ref{Freiden like equation}) would be equivalent.
When we define $\widetilde{{\cal Q}}$ as
\begin{eqnarray}
\label{structure Q} \widetilde{{\cal Q}} = \int \, d^{N}{\bf x} \,
P(\Theta) \, \left( \texttt{t\!F} - \sum_{n=1}^{N} \frac{\partial
ln P}{\partial \theta_{n}} ( \hat{\theta}_{n} - \theta_{n} )
\right) \,
\end{eqnarray}
then we obtain the structural equation of the IP form:
\begin{eqnarray}
\label{structure eq} &-& \widetilde{{\cal Q}} = \widetilde{{\cal
I}}
\\ &\equiv& \! \frac{1}{2} \! \int \!\! d^{N}{\bf x} \ P(\Theta) \!\!\!\! \sum_{n,
n'=1}^{N} \!\!\! ( - \frac{\partial^{2} ln P}{\partial \theta_{n'}
\partial \theta_{n}} ) ( \hat{\theta}_{n} - \theta_{n} ) (
\hat{\theta}_{n'} - \theta_{n'} ) . \nonumber
\end{eqnarray}
Now, let us postulate the validity of Eq.(\ref{structure eq}) on
the microscopic level:
\begin{eqnarray}
\label{micro structure eq}  & & \!\!\! \! \Delta_{LHS} \equiv
\sum_{n=1}^{N} 2 \ \frac{\partial ln P}{\partial \theta_{n}} (
\hat{\theta}_{n} - \theta_{n} ) - \sum_{n=1}^{N} 2 \
\frac{\texttt{t\!F}}{N} \nonumber \\ &=& \sum_{n,n'=1}^{N}
\texttt{i\!F}_{n n'} \, ( \hat{\theta}_{n} - \theta_{n} ) (
\hat{\theta}_{n'} - \theta_{n'} ) \equiv \Delta_{RHS} \; .
\end{eqnarray}
As the inverse of the covariance matrix the (observed) Fisher
information matrix:
\begin{eqnarray}
\label{observed IF} \texttt{i\!F} = \left(- \frac{\partial^{2} ln
P(\Theta)}{\partial \theta_{n'}
\partial \theta_{n}}\right) \,
\end{eqnarray}
is symmetric and positively defined. It follows that there is an
orthogonal matrix $U$ such that $\Delta_{RHS}$ in Eq.(\ref{micro
structure eq}) and hence $\Delta_{LHS}$ also, could be written in
the normal form:
\begin{eqnarray}
\label{normal form} \; \sum_{n,n'=1}^{N} &\texttt{i\!F}_{n n'}& \,
( \hat{\theta}_{n} - \theta_{n} ) ( \hat{\theta}_{n'} -
\theta_{n'} ) \nonumber \\ &\equiv& \Delta_{RHS} = \Delta_{LHS} =
\sum_{n=1}^{N} m_{n} \hat{\upsilon}_{n}^{2} \, ,
\end{eqnarray}
where $\hat{\upsilon}_{n}$ are some functions of
$\hat{\theta}_{n}$-s and $m_{n}$ are the elements (obtained for
$\Delta_{LHS}$) of the positive diagonal matrix $\texttt{m\!F}$
which because of the equality (\ref{normal form}) has to be equal
to the diagonal matrix obtained for $\Delta_{RHS}$, i.e.:
\begin{eqnarray}
\label{form of M} \texttt{m\!F} = D^{T} \ U^{T} \ \texttt{i\!F} \
U \ D \;\; .
\end{eqnarray}
Here $D$ is the scaling diagonal matrix with the elements $d_{n}
\equiv \sqrt{\frac{m_{n}}{\lambda_{n}}}$ and $\lambda_{n}$-s are
the eigenvalues of $\texttt{i\!F}$. Finally, we could rewrite
Eq.(\ref{form of M}) in the form of the Frieden structural
microscopic IP:
\begin{eqnarray}
\label{micro form of information eq} \texttt{q\!F} +
\texttt{i\!F} = 0 \; ,
\end{eqnarray}
where
\begin{eqnarray}
\label{micro form of qF} \texttt{q\!F} = - U \ (D^{T})^{-1} \
\texttt{m\!F} \ D^{-1} \ U^{T} \, .
\end{eqnarray}
The existence of the normal form (\ref{normal form}) is a very
strong condition which makes the whole Frieden analysis possible.
Two particular assumptions lead to the simple physical cases. When
the structure $\texttt{t\!F} = 0$ then from Eq.(\ref{micro
structure eq}) we obtain a form of the "master equation" (compare
with Eq.(\ref{L expand microscop})) with:
\begin{eqnarray}
\label{M for T zero}  \texttt{m\!F} = diag \left( 2\,
\frac{\partial ln P}{\partial \theta_{n}} \right) \, , \;\;
\hat{\upsilon}_{n} =  \sqrt{ \hat{\theta}_{n} - \theta_{n} } \, ,
\;\; \texttt{t\!F} = 0
\end{eqnarray}
and $ d_{n} = \sqrt{2 \, \frac{\partial ln P}{\partial
\theta_{n}}/\lambda_{n}}$. If we instead suppose that the
distribution is {\it regular} \cite{Pawitan} then with
$\frac{\partial ln P}{\partial \theta_{n}} = 0$ for all $n =
1,...,N$, we see from Eq.(\ref{micro structure eq}) that ($d_{n} =
\sqrt{2/\lambda_{n}}$):
\begin{eqnarray}
\label{M for logL zero}  \texttt{m\!F} =  (2 \, \delta_{n n'}) \;
,  \;\;\;  \hat{\upsilon}_{n} =  \sqrt{ \frac{\texttt{t\!F}}{N} }
\; .
\end{eqnarray}
After integrating Eq.(\ref{micro form of information eq}) with the
measure $d^{N}{\bf x} \, P(\Theta)$ we recover the integral
structural IP (postulated previously in \cite{Frieden} although in
a different form and interpretation but) in exactly the same form
as in \cite{Syska}:
\begin{eqnarray}
\label{expected form of information eq} Q + I = 0 \; ,
\end{eqnarray}
where $I$ is the Fisher information channel capacity:
\begin{eqnarray}
\label{iF and I} I = \int d^{N}{\bf x} \, P(\Theta) \;
\sum_{n,n'=1}^{N} (\texttt{i\!F})_{n n'} \;
\end{eqnarray}
and $Q$ is the $SI$:
\begin{eqnarray}
\label{qF and Q} Q = \int d^{N}{\bf x} \, P(\Theta) \;
\sum_{n,n'=1}^{N} (\texttt{q\!F})_{n n'} \; .
\end{eqnarray}
The IP given by Eq.(\ref{expected form of information eq}) is the
structural equation of many current physical models.

\section{The information transfer}
\label{information transfer}

As $I$ is the infinitesimal type of the Kulback-Leibler entropy
\cite{Frieden} which in the statistical estimation is used as a
tool in the model choosing procedure, hence the conjecture appears
that $I$ could be (after imposing the variational and structural
IPs) the cornerstone of the equation of motion (or generating
equation) of the physical system. These equations are to be the
best from the point of view of the IPs what is the essence of the
Frieden's EPI. The inner statistical thought in the Frieden method
is that the probing (sampling) of the space-time by the system
(even when not subjected to the real measurement) is performed by
the system alone, that using its proper field (and connected
amplitudes) of rank $N$, which is the size of the sample, probes
with its kinematical "Fisherian" degrees of freedom the position
space accessible for it. The transition from the statistical form
(\ref{Fisher_information_1}) of the $FI$ to its kinematical
representations (\ref{Fisher_information}) is given in
Footnote~2. \\
Let us consider the following informational scheme of the system.
Before the measurement takes place the system has $I$ of the
system which is contained in the kinematical degrees of freedom
and
%the $SI$ (structural information)
$Q$ of the system contained in the structural degrees of freedom,
as in Figure~1a.
\begin{figure*}[htb]%
%\subfloat[]
%{%
\includegraphics*[width=.47\textwidth,height=2.1cm]{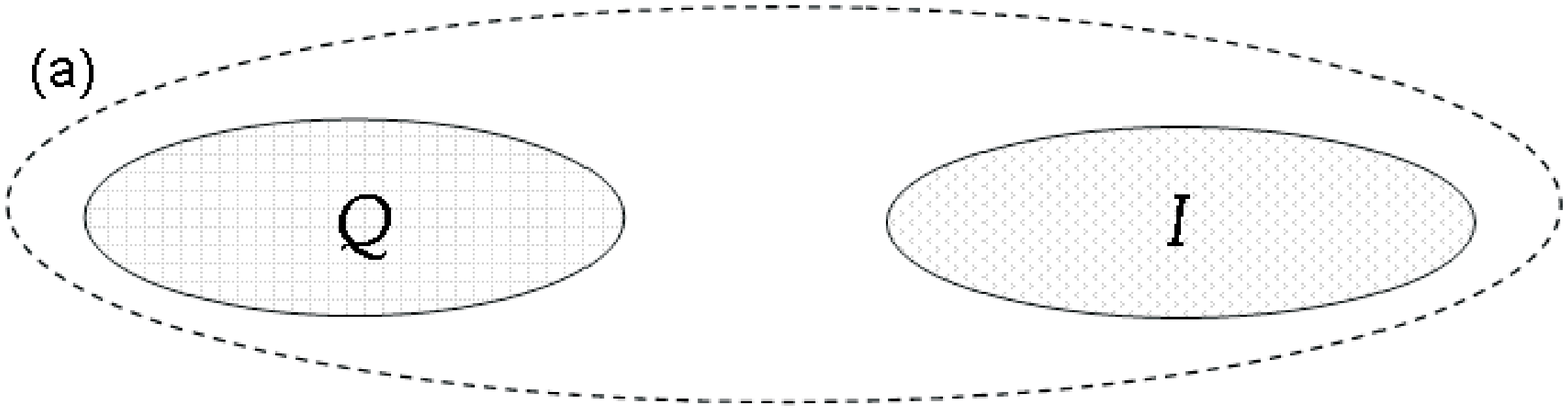}
%}
\hfill
%\subfloat[]{%
\includegraphics*[width=.47\textwidth,height=2.1cm]{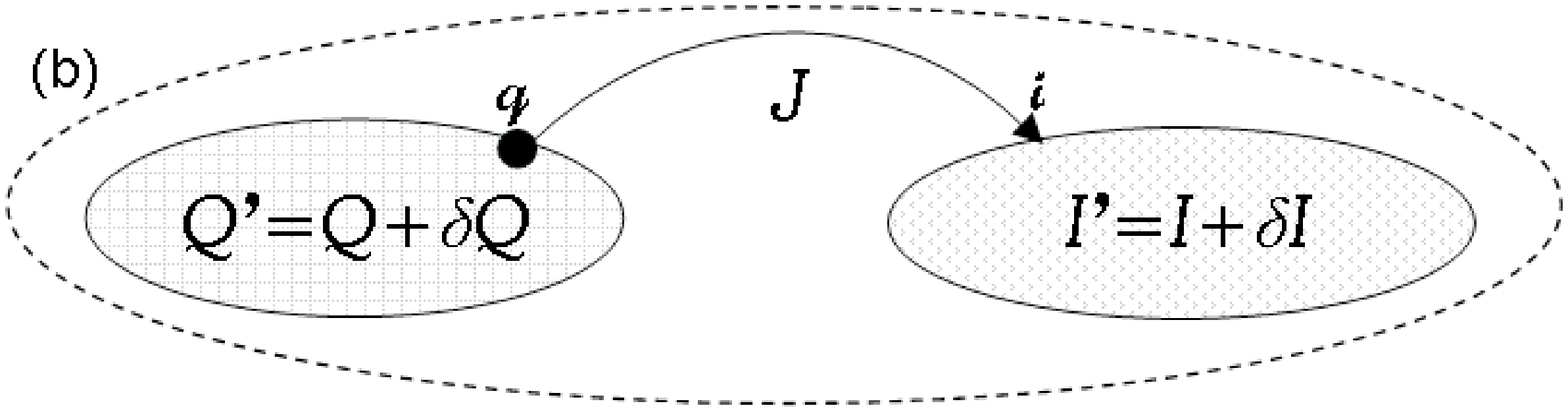}
%}%
\caption{%
Panel: (a)~The system before the measurement: $Q$ is the $SI$ of
the system contained in the structural degrees of freedom and $I$
is the $FI$ of the system contained in the kinematical degrees of
freedom. (b)~The system after the measurement: $Q'$ is the $SI$
and $I'$ is the $FI$ of the system; $\delta Q = Q' - Q \leq 0$ and
$\delta I = I' - I \geq 0$ as the transfer of the information
($TI$) in the measurement takes place with $J \geq 0$. In the
ideal measurement $\delta I = - \delta Q$.} \label{twosubfigures}
\end{figure*}
Now, let us "switch on" the measurement during which the {\it
transfer of the information} ($TI$) follows the rules~(see
Figure~1b):
\begin{eqnarray}
\label{delta Q and I} J \geq 0 \;\; {\rm hence} \;\; \delta I = I'
\! - I \geq 0 \; , \; \delta Q = Q' \! - Q \leq 0  ,
\end{eqnarray}
where $I'$, $Q'$ are the $FI$ and $SI$ after the measurement,
respectively and $J$ is the $TI$. We postulate that in the
measurement the $TI$ is ideal at the "point" $\emph{q}$, i.e. we
have that $Q = Q' + J$ $= Q + \delta Q + J$ hence  $\delta Q = -
J$. This means that the whole change of the $SI$ at the "point"
$\emph{q}$ is transferred. On the other hand at the "point"
$\emph{i}$ the rule for the $TI$ is that $I' \leq I + J$ hence $0
\leq \delta I = I' - I \leq J$. Therefore, as $J \geq 0$ we obtain
that $ |\delta I| \leq |\delta Q|$, the result which is sensible
as in the measurement information might be lost.~In the ideal
measurement we have obtained that $\delta Q = - \delta I$. \\
Previously we postulated the existence of the additive total
physical information ($TPI$) \cite{Syska}:
\begin{eqnarray}
\label{physical K} K = Q + I \; .
\end{eqnarray}
In \cite{Mroziakiewicz} the intuitive condition that $K \geq 0$
was chosen leading to the following form of the structural IP:
\begin{eqnarray}
\label{condition from K} \kappa \, Q + I = 0 \;
\end{eqnarray}
or
\begin{eqnarray}
\label{ideal condition from K} Q + I = 0 \;\;\; {\rm for} \;\;\;
\kappa = 1 \; ,
\end{eqnarray}
which we now derived in Eq.(\ref{expected form of information
eq}). In the case of Eq.(\ref{ideal condition from K}) we obtain
that $K$ is equal to
\begin{eqnarray}
\label{ideal K} K = Q + I = 0 \;\;\; {\rm for} \;\;\; \kappa = 1
\; .
\end{eqnarray}
The structural (internal) IP in Eq.(\ref{condition from K})
\cite{Syska} is operationally equivalent to the one postulated by
Frieden \cite{Frieden}, hence it has at least the same predictive
power. For the short description of the difference between both
approaches see \cite{Syska}. The other IP is the scalar
(variational) one. It has the form \cite{Syska}:
\begin{eqnarray}
\label{var K} \delta K = \delta (Q + I) = 0 \;\, \Rightarrow \;\,
K = Q + I \;\; {\rm is \; extremal}
\end{eqnarray}
The principles (\ref{condition from K}) and (\ref{var K}) are the
cornerstone of the Frieden EPI. They form the set of two
differential equations for the amplitudes $q_{n}$ which could be
consistent, leading for $\kappa = 1$ or $1/2$ to many well known
models of the field theory and statistical physics \cite{Frieden},
as we mentioned in the Introduction. It could be instructive to
recalculate them again using the new interpretation of $K$
\cite{Mroziakiewicz}. \\
Finally, let us notice that in the concorde with the postulated
behavior of the system in the measurement, we have obtained
$\delta I \leq J = - \delta Q$ from which it follows that:
\begin{eqnarray}
\label{K prim} K' = I' + Q' & \leq & (I + J) + (Q - J) = I + Q = K
\nonumber \\ & \Rightarrow &  \;\;\; K' \leq K \; .
\end{eqnarray}
In the case of $\delta I = - \delta Q$ we obtain $K' = K$. Then it
means that the $TPI$ remains unchanged in the ideal measurement
($\delta I = - \delta Q$) which, if performed on the intrinsic
level of sampling the space by the system alone (and not by the
observer), could possibly lead to the variational IP given by
Eq.(\ref{var K}).

\section{Conclusions}

The overwhelming impression of the EPI is that the $TPI$ is the
ancestor of the Lagrangian \cite{Frieden}. As the Fisherian
statistical formalism and hence the $IP$s lie here really as the
background of the description of the physical system it is not the
matter of interpretation only especially that the structural $IP$
(which is very close to the approach used in
\cite{Syska,Mroziakiewicz}) was derived in Section~\ref{structural
principle}. In the result, we have proved that the whole work of
Frieden with our equation of motion (or generating equation)
(\ref{structure eq}) lie inside the same EPI type of the
statistical modification of LM. In the contrary the master
equation (\ref{L expand}) lies in the other kind of the classical
statistics estimation, closer to the stochastic processes, than
the EPI method. Besides, for the exponential models with quadratic
($k=2$) Taylor expansion, the simple microscopic forms (\ref{M for
T zero}) and (\ref{M for logL zero}) are also given in
Section~\ref{structural principle}.  Yet e.g. for the family of
the distributions with $k
> 2$ and for the mixture family models \cite{Amari Nagaoka book}
the further investigations are needed. Here the alternative way of
solving the $IPs$ equations (\ref{condition from K}) and (\ref{var
K}) given by Frieden seems more appealing, as besides the boundary
conditions,
it is not restricted to the particular shape of the distributions. \\
The presented version of the EPI also has, as the original one
\cite{Frieden} has, the ability of the synonymous description of
the unmeasured and measured system, but additionally it is
characterized by the more precise distinction of these two
situations \cite{Mroziakiewicz}. Is is connected with the meaning
of $K$ as the $TPI$ which enables the precise discrimination of
the inner "measurement" performed by the system alone from the one
performed by an observer. Next, the chosen form of the internal IP
(\ref{condition from K}), suggests entanglement of the
observational space of the data with the whole unobserved space of
positions of the system \cite{Mroziakiewicz}. This situation is
also a little bit different than in \cite{Frieden} where (also
because of the interpretation of $K$) the entanglement with the
part of the unobserved configuration only was seen. E.g. in the
description of the EPR-Bohm effect our approach describes the
entanglement which takes place between the projection of the spin
of the observed particle and the unobserved joint configuration of
the system \cite{Mroziakiewicz}. Yet, the shared feature is that
$Q$\footnote{E.g. in \cite{Frieden} the scalar field case is
analyzed  for which $Q$ is connected with the rest mass of the
particle.} does not represent the $SI$ of the system only but the
information on the entanglement seen in the correlations of the
data in the measurement space also. Therefore in general the EPI
could be used as the tool in the estimation of the entangled
states. Our result is that the source of the (self)entanglement
could be understood as the consequence of the partition of the
Taylor expansion of the log-likelihood.  Now, for further
development of the EPI the differential geometry language
\cite{Amari Nagaoka book} of the $IPs$ is needed. Finally, let us
at least mention that there are other fields of science were EPI
could be used, e.g. the econophysics \cite{Frieden,PS1,PS2}. We
envisage that the formalism developed in this paper would be
applied in quantum information processing and quantum game theory
also \cite{PS1,PS2}.

\begin{acknowledgement}
This work is supported by L.J.CH..\\
This work  was supported in part by the scientific network {\it
Laboratorium Fizycznych Podstaw Przetwarzania Informacji}
sponsored by the {\it Polish Ministry of Science and Higher
Education} (decision no 106/E-345/BWSN-0166/2008) and by the
University of Silesia grant.
\end{acknowledgement}

% Use the following code if you wish to generate your bibliography with BibTeX;
% replace the string "pss-demo" below with the name(s) of
% the BibTeX data base(s) you want to use.
% The resulting bibliography-output (the contents of the .bbl file)
% must be pasted back into this file before submission.
%
% \bibliographystyle{pss}
% \bibliography{pss-demo}
%
% Replace the following example bibliography with your references
% before submission:

\end{document}